\documentclass[12pt]{article}
\usepackage{amsmath,amssymb,exscale}  
\input epsf

\def\gappeq{\mathrel{\rlap {\raise.5ex\hbox{$>$}}
{\lower.5ex\hbox{$\sim$}}}}
\def\permil{$\%\raise.20ex\hbox{$_0$}}
\def\lappeq{\mathrel{\rlap{\raise.5ex\hbox{$<$}}
{\lower.5ex\hbox{$\sim$}}}}

\begin{document}
\topmargin -1.0cm
\oddsidemargin -0.8cm
\evensidemargin -0.8cm
\pagestyle{empty}
\begin{flushright}
UNIL-IPT-00-14\\
hep-th/0006251\\
June 2000
\end{flushright}
\vspace*{5mm}

\begin{center}

{\Large\bf Living Inside a Hedgehog: Higher-dimensional
\vskip 0.1cm
Solutions that Localize Gravity}\\

\vspace{1.0cm}

{\large Tony Gherghetta\footnote{Email: tony.gherghetta@ipt.unil.ch}, 
Ewald Roessl\footnote{Email: ewald.roessl@ipt.unil.ch}
and Mikhail Shaposhnikov\footnote{Email: mikhail.shaposhnikov@ipt.unil.ch}}\\
\vspace{.6cm}
{\it {Institute of Theoretical Physics\\ University of Lausanne\\ 
CH-1015 Lausanne, Switzerland}}
\vspace{.4cm}
\end{center}

\vspace{1cm}  
\begin{abstract}  

We consider spherically symmetric higher-dimensional solutions of
Einstein's equations with a bulk cosmological constant and  $n$
transverse dimensions.  In contrast to the case of one or two extra
dimensions we find no solutions that localize gravity when  $n \geq
3$, for strictly local topological defects. We discuss global
topological defects that lead to the localization of gravity and 
estimate the corrections to Newton's law. We show that the introduction 
of a bulk ``hedgehog'' magnetic field  leads to a regular geometry and
localizes gravity on the 3-brane with either a positive, zero 
or negative bulk cosmological constant. The  corrections to Newton's 
law on the 3-brane are  parametrically the same as for the case of one
transverse dimension.

\end{abstract}

\vfill

\eject
\pagestyle{empty}
\setcounter{page}{1}
\setcounter{footnote}{0}
\pagestyle{plain}


\section{Introduction} 

A lot of attention has been devoted recently to alternatives
\cite{RS2}--\cite{rusu} of Kaluza-Klein compactification~\cite{KK}.
In particular, our spacetime can be associated with some topological
defect - 3 brane, embedded in a higher-dimensional spacetime with
non-compact extra dimensions. It is usually assumed that the matter
fields are localized on the brane because of the specific dynamics of
solitons in string theory - D-branes~\cite{branes}. Moreover, in Ref.
\cite{rusu} it was shown that the gravity of a domain wall in
5-dimensional anti-deSitter (AdS) spacetime has a 4-dimensional 
character for the
particles living on the brane, provided that the domain wall 
tension is fine tuned to a bulk cosmological constant. The corrections 
to Newton's gravity law are generically small for macroscopic scales (see,
however, \cite{greg} for a more complicated construction involving
several branes). A similar statement is true for a local string
living in 6-dimensional AdS space~\cite{gs} 
(or more general constructions~\cite{cp}).

The aim of the present paper is to generalize the results of 
Ref.~\cite{gs} to the case when the number $n$ of transverse 
dimensions is larger than two. This happens to be not as trivial  as
one expects. The reason is that the transverse spaces with   $n\leq
2$, and $n \geq 3$ extra dimensions are qualitatively different, at
least in the spherically symmetric setup we are interested in. In
contrast to the case with $n \geq 3$, for  $n=1$ the extra space is
flat while for $n=2$ the extra space can be curved, but is still
conformally flat. In the absence of a brane the possibilities of
compactification were studied in \cite{RS1} for $n=2$ and in
\cite{ran} for general $n$.

We will consider three different possibilities. The first possibility
is called a strictly local defect. By strictly local we mean the
situation when the stress-energy tensor of the defect is zero outside
the core (or, for the more realistic situation of a ``fat'' brane, 
exponentially falling outside the core). 
Here, we were not able to find any geometry leading
to the compactification of gravity, contrary to the $n=1$ and $n=2$
cases. 

The second possibility is related to the so called global defects.
In this case one assumes that there exists a scalar field with, say $O(N)$ 
($N\geq n$) global symmetry which is spontaneously broken. Outside the
string core this field may have a  hedgehog type configuration, which
gives a specific contribution to the energy-momentum tensor outside
the defect. This case was studied in \cite{ck,rg} for $n=2$ and in
\cite{ov} for higher dimensions, where the solutions with an
exponential warp factor were found. We compute the corrections to
Newton's law in this case and study the boundary conditions at the core
of the global string. Furthermore, a generalization
of these metric solutions is also defined.

The third  possibility is related to configurations of the monopole
type, where outside the defect there exists a magnetic field (or, its
generalization to higher dimensions - $p$-form Abelian gauge field). We
consider different spherically symmetric ansatz and define those that
lead to gravity localization on the 3-brane and a
regular geometry in the bulk. We also discuss the corrections to
Newton's law for these solutions. In the absence of a brane a model
with an abelian magnetic field in six dimensions was considered in
\cite{gib}.

\section{Einstein equations with a 3-brane source}

In D-dimensions the Einstein equations with a bulk cosmological
constant $\Lambda_D$  and stress-energy tensor $T_{AB}$ are
\begin{equation}
    R_{AB} - \frac{1}{2} g_{AB} R = \frac{1}{M_D^{n+2}}\left(\Lambda_D
   g_{AB} + T_{AB}\right)~,
\end{equation}
where $M_D$ is the reduced D-dimensional Planck scale.  We will
assume  that there exists a solution that respects 4d Poincare
invariance.  A D-dimensional metric satisfying this ansatz for $n$
transverse spherical coordinates with $0\leq \rho < \infty$,  
$0\leq \{\theta_{n-1},\dots,\theta_2\} < \pi$ and $0\leq \theta_1
< 2\pi$, is
\begin{equation}
\label{metric}
    ds^2 = \sigma(\rho) g_{\mu\nu} dx^\mu dx^\nu 
    -d\rho^2-\gamma(\rho)d\Omega_{n-1}^2\, ,
\end{equation}
where  the metric signature of $g_{\mu\nu}$ is $(+,-,-,-)$ 
and $d\Omega_{n-1}^2$ is defined recursively as
\begin{equation}
     d\Omega_{n-1}^2 = d\theta_{n-1}^2 + 
     \sin^2\theta_{n-1} d\Omega_{n-2}^2~,
\end{equation}
with $d\Omega_0^2 = 0$. 
At the origin $\rho=0$ we will assume that there is a 3-brane, whose
source is described by a stress-energy tensor $T^A_B$ with nonzero
components
\begin{equation}
\label{source}
     T^\mu_\nu = \delta^\mu_\nu f_0(\rho), \quad T_\rho^\rho = f_\rho(\rho),
     \quad {\rm and} \quad T_\theta^\theta = f_\theta(\rho)~.
\end{equation}
Here we have introduced three source functions $f_0$, $f_\rho$, and
$f_\theta$ which depend only on the radial coordinate $\rho$ and by
spherical symmetry  all the angular source functions are identical, where
we have defined $\theta\equiv \theta_{n-1}$. 
Using the metric ansatz  (\ref{metric}) and the stress-energy tensor
(\ref{source}), the Einstein equations  become 
\begin{eqnarray}
\label{solnset1}
    \frac{3}{2} \frac{\sigma^{\prime\prime}} {\sigma} 
   +\frac{3}{4}(n-1)\frac{\sigma^\prime}{\sigma}\frac{\gamma^\prime}{\gamma}
   +\frac{1}{8}(n-1)(n-4) \frac{\gamma^{\prime 2}}{\gamma^2}
   +\frac{1}{2}(n-1)\frac{\gamma^{\prime\prime}}{\gamma}
   -\frac{1}{2\gamma}(n-1)(n-2) \nonumber\\
    = -\frac{1}{M_D^{n+2}}(\Lambda_D + f_0(\rho)) +
    \frac{\Lambda_{phys}}{M_P^2}\frac{1}{\sigma}~,
    \\
    \frac{3}{2}\frac{\sigma^{\prime 2}}{\sigma^2}
   +(n-1)\frac{\sigma^\prime}{\sigma}\frac{\gamma^\prime}{\gamma}
   +\frac{1}{8}(n-1)(n-2) \frac{\gamma^{\prime 2}}{\gamma^2}
   -\frac{1}{2\gamma}(n-1)(n-2) \nonumber\\
    = -\frac{1}{M_D^{n+2}}(\Lambda_D + f_\rho(\rho)) +
    \frac{2\Lambda_{phys}}{M_P^2}\frac{1}{\sigma}~,
    \\
\label{solnset3}
    2\frac{\sigma^{\prime\prime}} {\sigma} 
    +\frac{1}{2}\frac{\sigma^{\prime 2}}{\sigma^2}
   +(n-2)\frac{\sigma^\prime}{\sigma}\frac{\gamma^\prime}{\gamma}
   +\frac{1}{8}(n-2)(n-5) \frac{\gamma^{\prime 2}}{\gamma^2}
   +\frac{1}{2}(n-2)\frac{\gamma^{\prime\prime}}{\gamma}
   -\frac{1}{2\gamma}(n-2)(n-3) \nonumber\\
    = -\frac{1}{M_D^{n+2}}(\Lambda_D + f_\theta(\rho)) 
    +\frac{2\Lambda_{phys}}{M_P^2}\frac{1}{\sigma}~,
\end{eqnarray}
where the $^\prime$ denotes differentiation $d/d\rho$ and the Einstein
equations arising from all the angular components simply reduce to the
one angular equation (\ref{solnset3}). The constant
$\Lambda_{phys}$ represents the physical 4-dimensional
cosmological constant, where
\begin{equation}
     R_{\mu\nu}^{(4)} - \frac{1}{2} g_{\mu\nu} R^{(4)} = 
     \frac{\Lambda_{phys}}{M_P^2} g_{\mu\nu}~.
\end{equation}
The system of equations (\ref{solnset1})--(\ref{solnset3}) for $f_i=0$
was first derived in \cite{ran} and describes
the  generalization of the setup considered in \cite{RS1,rusu,gs,ov}, to
the case  where there are $n$ transverse dimensions, together with a
nonzero cosmological constant in 4-dimensions and stress-energy tensor
in the bulk. If we eliminate two of
the equations in (\ref{solnset1})--(\ref{solnset3}) then the source
functions satisfy
\begin{equation}
     f_\rho^\prime = 2 \frac{\sigma'}{\sigma} (f_0-f_\rho) 
      + \frac{n-1}{2} \frac{\gamma'}{\gamma} (f_\theta-f_\rho)~,
\end{equation}
which is simply a consequence of the conservation of the stress-energy
tensor $D_M T^M_N = 0$. In general the Ricci scalar corresponding to 
the metric ansatz (\ref{metric}) is
\begin{eqnarray}
\label{ricci}
      R=4\frac{\sigma''}{\sigma}+ \frac{\sigma^{\prime2}}{\sigma^2}+
       2(n-1) \frac{\sigma'}{\sigma}\frac{\gamma'}{\gamma}  + (n-1)
       \frac{\gamma''}{\gamma}+
       \frac{1}{4}(n-1)(n-4)\frac{\gamma^{\prime2}}{\gamma^2} \nonumber\\
       - (n-1)(n-2)\frac{1}{\gamma} -
       \frac{4}{\sigma}\frac{\Lambda_{\rm phys}}{M_P^2}~.
\end{eqnarray}

The boundary conditions at the origin of the transverse space
are assumed to be
\begin{equation}
     \sigma^\prime\big|_{\rho=0} = 0~, \quad (\sqrt{\gamma})^\prime
     \big|_{\rho=0} = 1  \quad {\rm and} 
     \quad \gamma\big|_{\rho=0} = 0~,
\end{equation}
which is consistent with the usual regular solution in flat space.
We have set $\sigma(0) =A$, where $A$ is a constant.
Following~\cite{fiu}, we can integrate over the disk of small radius
$\epsilon$ containing the 3-brane, and define various components of 
the brane tension per unit length as
\begin{equation}
     \mu_i = \int_0^\epsilon d\rho\, \sigma^2 \gamma^{(n-1)/2}\, 
     f_i(\rho)~.
\end{equation}
where $i = 0,\rho,\theta$. Using the system of equations
(\ref{solnset1})--(\ref{solnset3}) we obtain the following boundary 
conditions 
\begin{equation}
\label{junc1}
     \sigma\sigma^\prime\sqrt{\gamma^{n-1}} \big|_0^\epsilon = 
       \frac{2}{(n+2)}\frac{1}{M_D^{n+2}}\bigg((n-2)\mu_0-\mu_\rho 
       -(n-1)\mu_\theta\bigg)~,
\end{equation}
and
\begin{equation}
\label{junc2}
     \sigma^2\sqrt{\gamma^{n-2}}(\sqrt{\gamma})^\prime \big|_0^\epsilon = 
       -\frac{1}{(n+2)}\frac{1}{M_D^{n+2}}\bigg(4\mu_0+\mu_\rho
     -3\mu_\theta\bigg)~,
\end{equation}
where it is understood that the limit $\epsilon\rightarrow 0$ is
taken. The equations (\ref{junc1}) and (\ref{junc2}) are the general
conditions relating the brane tension components to the metric
solution of the Einstein equations
(\ref{solnset1})--(\ref{solnset3}),  and lead to nontrivial
relationships between the  components of the brane tension per unit
length. In particular, these conditions on the brane tension
components reduce to the  relations obtained for $n=2$~\cite{gs}.
Furthermore, by analogy with the solution  for local strings we can
identify (\ref{junc1}) as the gravitational mass  per unit length and
(\ref{junc2}) as the angular deficit per unit length. Thus the source
for the 3-brane, in general curves the transverse space.

From the Einstein term in the D-dimensional Lagrangian we can obtain
the  effective four-dimensional Planck mass. Using the spherically
symmetric  metric ansatz (\ref{metric}), the four-dimensional reduced
Planck mass is given by
\begin{equation}
        M_P^2 = {\cal A}_n M_D^{n+2} \int_0^\infty d\rho\, \sigma\, 
        \gamma^{(n-1)/2}~.
\end{equation}
where ${\cal A}_n$ is the surface area of an $n$-dimensional unit
sphere. We are interested in obtaining solutions to the Einstein
equations  (\ref{solnset1})--(\ref{solnset3}) such that a finite
four-dimensional Planck mass is obtained. This leads to various
possible asymptotic behaviours for the metric warp factors $\sigma$
and $\gamma$ in the  limit $\rho\rightarrow\infty$. Below, we will
concentrate only on the case when the 4-dimensional cosmological constant
is zero, $\Lambda_{phys}=0$.

\section{Strictly local defect solutions}

First, let us assume that the functions $f_i(\rho)$ are zero outside
the core of the topological defect. In order to obtain a finite
4-dimensional Planck scale, one requires a solution of the system of
equations  (\ref{solnset1})--(\ref{solnset3}) for which  the function
$\sigma \gamma^{(n-1)/2}$ goes to zero when  $\rho\rightarrow\infty$.
For $n=1$ and $n=2$ the solutions are known to exist, see \cite{rusu}
and \cite{gs} correspondingly. However, when $n \geq 3$ the structure
of the equations is qualitatively different, because now there is a
$1/\gamma$ term. Thus, there is no simple generalization of the
solutions found for $n=1,2$.

To neutralize the effect of the $1/\gamma$ term, one can look for
asymptotic solutions for which $\gamma$ is a positive constant.  However, one
can easily check that the system of equations 
(\ref{solnset1})--(\ref{solnset3}) does not allow a solution for
which $\gamma$ tends to a constant when $\rho \rightarrow \infty$,
and $\sigma$ is a negative exponential.  

Alternatively, we can assume that there is an asymptotic solution for
which $\gamma \rightarrow \infty$ but $\sigma$ tends to zero faster
than $\gamma^{(n-1)/2}$ and omit the troublesome $1/\gamma$ term from
the equations of motion. In this case  the set of
equations~(\ref{solnset1})--(\ref{solnset3}) can  be simply reduced
to a single equation, as in 6D case \cite{RS1,gs}:
\begin{equation}
\label{zeqn}
     z'' = -\frac{d U(z)}{dz}~,
\end{equation}
where the potential $U(z)$ is given by
\begin{equation}
\label{upot}
     U(z) = \frac{(n+3)}{4(n+2)}\,\frac{\Lambda_D}{M_D^{n+2}}\, z^2~.
\end{equation}
With this parametrisation the metric functions $\sigma(\rho)$ and
$\gamma(\rho)$  can be written in terms of $z(\rho)$ as
\begin{eqnarray}
     \sigma &=& |z'|^{(2-\sqrt{(n+2)(n-1)})/(n+3)}\,
        |z|^{(2+\sqrt{(n+2)(n-1)})/(n+3)} \\
     \gamma &=& |z'|^{6/(1-n+2\sqrt{(n+2)(n-1)})}\,
        |z|^{6/(1-n-2\sqrt{(n+2)(n-1)})}~.
\end{eqnarray}
Solving equation (\ref{zeqn}) with the potential (\ref{upot}) gives 
the general solution
\begin{equation}
\label{zsoln}
           z(\rho) = d_1\, e^{-\frac{1}{4}(n+3) c \rho} + d_2\, 
           e^{\frac{1}{4}(n+3) c \rho}~,
\end{equation}
where $c^2=-8\Lambda_D/((n+2)(n+3) M_D^{n+2})$, 
$d_1, d_2$ are constants and we take $\Lambda_D<0$. This
solution is a generalization of the pure exponential solution
considered earlier and in Ref.~\cite{ov}, see below.  In this picture
we can think of particle motion under the influence of the potential
(\ref{upot}) with position $z(\rho)$ and ``time'' $\rho$. 

Since $1-n+2\sqrt{(n+2)(n-1)}>0$ and $1-n-2\sqrt{(n+2)(n-1)} < 0$,
the metric factor $\gamma$ can be large in two cases. In the  first
case $z(\rho)$ is zero for some $\rho_0$. However, this point is only
a coordinate singularity (the Ricci scalar is regular  at this point)
and the metric can be extended beyond $\rho_0$, leading  then to an
exponentially rising solution for both $\sigma$ and $\gamma$. This,
unfortunately, is not interesting for compactification.  In the
second case both $z'$ and $z$ are non-zero and increase exponentially
for large $\rho$. Thus, there is no possibility of a finite Planck
scale in this case either.

Similarly, we were unable to find solutions in the  reverse case when
$\gamma$ vanishes at infinity. Moreover, even if such solutions were
to exist, they would likely lead to  a singular geometry (naked
singularity), because the Ricci scalar contains a $1/\gamma$ term,
see eq. (\ref{ricci}). This was indeed shown to be the case for solutions
with regular geometries at $\rho=0$ in \cite{ran}.

\section{Bulk scalar field}
\subsection{Global topological defects}

The other possibility is to consider defects with different types of
``hair'' i.e. with non-zero stress-energy tensor outside the core of
the defect.  We start with global topological defects. In fact, we
have little to add to this question as it has been extensively
studied in \cite{rg,ov}, so we just list a number of explicit
solutions (see also~\cite{dvali,oda,luty}).  
Again, for simplicity we will restrict to the case where
the four-dimensional cosmological constant $\Lambda_{phys} =0$, and
assume that outside the core $(\rho > \epsilon)$
\begin{equation}
\label{solnform}
     \sigma(\rho) = e^{-c \rho}~.
\end{equation}
We have chosen the arbitrary integration constant, which 
corresponds to an overall rescaling of the coordinates $x^\mu$,
such that $\displaystyle \lim_{\epsilon\rightarrow 0} 
\sigma(\epsilon)=1$.

Consider $n$ scalar fields $\phi^a$ with a potential
\begin{equation}
        V(\phi) = \lambda (\phi^a\phi^a - v^2)^2~,
\end{equation}
where $v$ has mass dimension $(n+2)/2$.
Then the potential minimum is at $\phi^a\phi^a = v^2$. 
The defect solution has a 
``hedgehog'' configuration outside the core
\begin{equation}
        \phi^a(\rho) = v d^a~,
\end{equation}
where $d^a$ is a unit vector in the extra dimensions, $d^n=
\cos\theta_{n-1},~d^{n-1} = \sin\theta_{n-1} \cos\theta_{n-2},\dots$.

The scalar field gives an additional contribution to the stress-energy 
tensor in the bulk with components
\begin{eqnarray}
         T_\mu^\nu &=& (n-1) \frac{v^2}{2\gamma} \delta_\mu^\nu~, \\
         T_\rho^\rho &=& (n-1) \frac{v^2}{2\gamma}~, \\
         T_\theta^\theta &=& (n-3) \frac{v^2}{2\gamma}~.
\end{eqnarray}
Now, for $\Lambda_D <0$ and $v^2 > (n-2) M_D^{n+2}$ the following
solution leads to the localization of gravity and a regular geometry in
the bulk~\cite{rg,ov}
\begin{eqnarray}
\label{s1}
        c&=&\sqrt{\frac{2(-\Lambda_D)}{(n+2) M_D^{n+2}}}~,\\
\label{s2}
      \gamma &=&\frac{1}{c^2}\left(\frac{v^2}{M_D^{n+2}}-n+2 
       \right)~,
\end{eqnarray}
provided that the brane tension components satisfy the conditions 
\begin{equation}
\label{junc11}
     -c\sqrt{\gamma^{n-1}} = 
       \frac{2}{(n+2)}\frac{1}{M_D^{n+2}}\bigg((n-2)\mu_0-\mu_\rho 
       -(n-1)\mu_\theta\bigg)~,
\end{equation}
and
\begin{equation}
\label{junc12}
     A^2 \delta_{n2} = 
       \frac{1}{(n+2)}\frac{1}{M_D^{n+2}}\bigg(4\mu_0+\mu_\rho
     -3\mu_\theta\bigg)~.
\end{equation}
The transverse geometry of this solution is that of a cylinder, 
$R_+ \times S_{n-1}$ with $R_+$ being the half-line and $S_{n-1}$ 
being an $n-1$-sphere. 

When $v^2 = (n-2)M_D^{n+2}$ the $1/\gamma$ terms are eliminated
from  the system of equations (\ref{solnset1})--(\ref{solnset3}) and
the exponential solution to the coupled set of equations 
(\ref{solnset1})--(\ref{solnset3}) can 
then be found with 
\begin{equation}
  \gamma(\rho)=R_0^2\sigma(\rho),~~
   c=\sqrt{\frac{8(-\Lambda_D)}{(n+2)(n+3) M_D^{n+2}}}~,
     \label{sgsol}
\end{equation}
where $R_0$ is an arbitrary length scale. As expected, the negative
exponential solution (\ref{solnform}) requires that $\Lambda_D < 0$.
Notice that the exponential solution (\ref{sgsol}) only requires the
``hedgehog'' scalar field configuration in the bulk for transverse
spaces with dimension $n\geq 3$.  No such configuration is needed for
the 5d~\cite{rusu} and 6d cases~\cite{gs},  which only require
gravity in the bulk.

The Ricci scalar corresponding to the negative exponential solution
with vanishing four-dimensional cosmological constant is 
\begin{equation}
      R=(n+3)(n+4)\frac{c^2}{4} -(n-1)(n-2)\frac{e^{c \rho}}{R_0^2}~,
\end{equation}
which diverges when $\rho=\infty$.  Thus we see that for $n\geq 3$
the space is no longer  a constant curvature space and in fact has a
singularity at $\rho=\infty$.  This is also confirmed by looking at
the other curvature invariants, $R_{AB}R^{AB}$ and
$R_{ABCD}R^{ABCD}$. Only for the 5d and 6d cases do we obtain a
constant  curvature anti-deSitter space. The appearance of a
singularity is similiar to the case of the global-string
defect~\cite{ck}. 

The metric solution (\ref{sgsol}) can also be written in the form
\begin{equation}
    ds^2 = z^2 g_{\mu\nu} dx^\mu dx^\nu - z^2 R_0^2 d\Omega_{n-1}^2
         -\frac{4}{c^2 z^2} dz^2~.
\end{equation}
where $z=\exp(-\frac{c}{2} \rho)$. In these coordinates 
the origin $\rho = 0$ is now mapped to $z=1$ and the 3-brane 
source is spread around the surface area of a $n$-dimensional
sphere of radius $R_0$. This confirms our previous suggestion~\cite{gs}
that for $n\geq 2$ transverse dimensions 
the 3-brane can be identified with ${\cal M}_{n+2}/S^{n-1}$, where
${\cal M}_{n+2}$ is a $(n+2)$-brane.

Requiring that our exponential solution satisfy the boundary conditions
(\ref{junc1}) and (\ref{junc2}), leads to the relation
\begin{equation}
        \mu_\theta+ \mu_\rho = \frac{1}{2} (n+2) R_0^{n-1} M_D^{n+2}c ~,
\end{equation}
where $\mu_0$ satisfies
\begin{equation}
     \mu_0 = \mu_\theta + A^2 M_D^4 \,\delta_{n2}~,
\end{equation}
and $\mu_\rho$ remains undetermined. Thus for $n>2$ we simply
have $\mu_0 = \mu_\theta$.

\subsection{Corrections to Newton's Law}

For the solution (\ref{s1}),(\ref{s2}) the corrections to Newton's law
are parametrically the same as for 5d case, since $\gamma$ is a constant.
On the other hand, the singular solution (\ref{sgsol})
will ultimately require that the singularity is smoothed by string 
theory corrections (perhaps similiar to the nonsingular deformations 
considered in~\cite{cpt}). Assuming that this is the case, then
the corrections to Newton's law on the 3-brane can be calculated by
generalizing the calculation presented in~\cite{rusu,gs} 
(see also~\cite{cehs}).

In order to see that gravity is only localized on the 3-brane, let us
now consider the equations of motion for the linearized metric
fluctuations. We will only concentrate on the spin-2 modes and
neglect the scalar modes, which needs to be taken into account
together with the bending of the brane~\cite{bend}. The vector modes
are massive as follows from a simple modification of the results in
Ref.~\cite{lt}. For a fluctuation of the form $h_{\mu\nu}(x,z) =
\Phi(z) h_{\mu\nu}(x)$ where $z=(\rho,\theta)$ and $\partial^2
h_{\mu\nu}(x)  = m_0^2 h_{\mu\nu}(x)$  we can separate the variables
by defining $\Phi(z)= \sum_{l_i m} \phi_m(\rho)
Y_{l_i}(\phi,\theta_i)$.  The radial modes satisfy the
equation~\cite{lt}
\begin{equation}
\label{diffop}
     -\frac{1}{\sigma \gamma^{(n-1)/2}}\,\partial_\rho\left[
      \sigma^2\gamma^{(n-1)/2} \,\partial_\rho \phi_m  \right] = 
      m^2 \phi_m~,
\end{equation}
where $m_0^2=m^2 + \Delta^2/R_0^2$ contains the contributions from the 
angular momentum modes $l_i$. The differential operator (\ref{diffop}) 
is self-adjoint provided that we impose the boundary conditions
\begin{equation}
\label{bc}
     \phi_m^\prime(0) = \phi_m^\prime(\infty) = 0~,
\end{equation}
where the modes $\phi_m$ satisfy the orthonormal condition
\begin{equation}
       {\cal A}_n \int_0^\infty d\rho \, \sigma \,\gamma^{(n-1)/2}\,
       \phi_m^\ast \phi_n = \delta_{mn}~.
\end{equation}
Using the specific solution
(\ref{solnform}),(\ref{sgsol}), the differential operator
({\ref{diffop}) becomes
\begin{equation}
\label{diffeqn}
      \phi_m^{\prime\prime} -\frac{(n+3)}{2}\, c\, \phi_m^\prime +
     m^2 e^{c\rho}  \phi_m = 0~.
\end{equation}
This equation is the same as that obtained for the 5d domain wall
solution~\cite{rusu}, when $n=1$ and the local stringlike
solution~\cite{gs} when $n=2$. We see that each extra transverse
coordinate augments this coefficient by $1/2$. When $m=0$ we clearly
see that $\phi_0(\rho)$ = constant is a solution. Thus we have a
zero-mode tensor fluctuation which is localized near  the origin
$\rho =0$ and is normalizable.

The contribution from the nonzero modes will modify Newton's law on the
3-brane. In order to calculate this contribution we need
to obtain the wavefunction for the nonzero modes at the origin. The 
nonzero mass eigenvalues can be obtained by imposing the boundary conditions
(\ref{bc}) on the solutions of the differential equation (\ref{diffeqn}). 
The  solutions of (\ref{diffeqn}) are
\begin{equation}
    \phi_m(\rho) = e^{\frac{c}{4} (n+3)\rho}
    \left[ C_1 J_{\frac{1}{2}(n+3)}(\frac{2m}{c} e^{\frac{c}{2}\rho}) 
        + C_2 Y_{\frac{1}{2}(n+3)}(\frac{2m}{c} e^{\frac{c}{2}\rho}) \right]~,
\end{equation}
where $C_1,C_2$ are constants and $J_{\frac{1}{2}(n+3)},
Y_{\frac{1}{2}(n+3)}$ are  the usual Bessel functions. Imposing the
boundary conditions (\ref{bc}) at a finite radial distance cutoff
$\rho=\rho_{\rm max}$ (instead of $\rho=\infty$) will lead to a
discrete mass spectrum, where for $k=1,2,3,\dots$ we obtain 
\begin{equation}
      m_k \simeq c(k+\frac{n}{4}) \frac{\pi}{2} e^{-\frac{c}{2} \rho_{\rm max}}~.
\end{equation}
With this discrete mass spectrum we find that
in the limit of vanishing mass $m_k$,
\begin{equation}
      \phi_{m_k}^2(0) = \frac{1}{{\cal A}_n R_0} \frac{\pi c}{8}
     \frac{(n+1)^2}{\Gamma^2[(n+3)/2]} \left(\frac{m_k}{c}\right)^n 
     e^{-\frac{c}{2} \rho_{\rm max}}~,
\end{equation}
where $\Gamma[x]$ is the gamma-function.
On the 3-brane the gravitational potential between two point masses
$m_1$ and $m_2$, will receive a contribution from the discrete nonzero modes 
given by
\begin{equation}
      \Delta V(r) \simeq G_N  \frac{m_1 m_2}{r}  \frac{\pi
      (n+1)}{4\Gamma^2[(n+3)/2]} \sum_k e^{-m_k r}
      \left(\frac{m_k}{c}\right)^n  e^{-\frac{c}{2} \rho_{\rm max}}~,
\end{equation}
where $G_N$ is Newton's constant.
In the limit that $\rho_{\rm max}\rightarrow \infty$, the spectrum
becomes continuous and the discrete sum is converted into an
integral. The contribution to the gravitational potential then
becomes
\begin{eqnarray}
       V(r)
       &\simeq& G_N \frac{m_1 m_2}{r} \left[1+\frac{1}{2 c^{n+1}}
       \frac{n+1}{\Gamma^2[(n+3)/2]} \int_0^\infty dm\,m^n e^{-m
       r}\right] \\
       &=& G_N \frac{m_1 m_2}{r} \left[1+ \frac{\Gamma[n+2]}{2
       \Gamma^2[(n+3)/2]} \frac{1}{(cr)^{n+1}}\right]
\end{eqnarray}
Thus we see that for $n$ transverse dimensions
the correction to Newton's law from the bulk continuum states grows 
like $1/(cr)^{n+1}$. This correction becomes more suppressed as the number
of transverse dimensions grows, because now the  gravitational field of
the bulk continuum modes spreads out in more dimensions and so
their effect on the 3-brane is weaker.

\section{Bulk $p$-form field}

The global topological defects considered in the previous section
inevitably contain massless scalar fields -- Nambu-Goldstone bosons 
associated with the spontaneous breakdown of the global symmetry. 
Thus, the stability of these configurations is far from being obvious.

We will now consider the possibility of introducing other types of
bulk fields ($p$-form field $A_{\mu_1\dots\mu_p}$), which directly
lead to a regular geometry. The stability of the corresponding
configurations  may be insured simply by the magnetic flux
conservation. The D-dimensional action is
\begin{equation}
     S= \int d^Dx \sqrt{|g|} \left( \frac{1}{2} M_D^{n+2} R - 
     \frac{\Lambda_D}{M_D^{n+2}} + (-1)^p\frac{1}{4} 
      F_{\mu_1\dots\mu_{p+1}}F^{\mu_1\dots\mu_{p+1}} \right)~.
\end{equation}
The energy-momentum tensor associated with the $p$-form field
configuration is given by
\begin{equation}
     T^A_B = (-1)^{p+1}\left( \frac{1}{4}\delta^A_B
      F_{\mu_1\dots\mu_{p+1}}F^{\mu_1\dots\mu_{p+1}} -
       \frac{p+1}{2} F^A_{\quad\mu_1\dots \mu_p} F_B^{\quad\mu_1\dots \mu_p}\right)~.
\end{equation}
A solution to the equation of motion for the $p$-form field when
$p=n-2$ is
\begin{equation}
      F_{\theta_1\dots\theta_{n-1}} = Q
(\sin\theta_{n-1})^{(n-2)}\dots \sin\theta_2~,
\end{equation}
where $Q$ is the charge of the field configuration and all other
components of $F$ are equal to zero. In fact, this ``hedgehog''
field configuration is the generalization of the magnetic field of a monopole.
The stress-energy tensor associated with this $p$-form field in the bulk is
\begin{eqnarray}
      T^\mu_\nu &=& \frac{(n-1)!}{4}\frac{Q^2}{\gamma^{n-1}}
      \delta^\mu_\nu~, \\
      T^\rho_\rho &=& -T^\theta_\theta =
        \frac{(n-1)!}{4}\frac{Q^2}{\gamma^{n-1}}~.
\end{eqnarray}
Let us assume a solution outside the core $(\rho>\epsilon)$
of the form
\begin{equation}
\label{pfansatz}
    \sigma(\rho)= e^{-c\rho} \quad {\rm and} \quad \gamma= {\rm constant}~,
\end{equation}
where we have again chosen the arbitrary integration constant such that
$\displaystyle \lim_{\epsilon\rightarrow 0} \sigma(\epsilon)=1$.
With this ansatz we see from the Ricci scalar that the transverse
space  will have a constant curvature and the effective
four-dimensional Planck constant will be finite.  If we substitute
this ansatz and also include the contribution of the $p$-form bulk
field to the stress-energy tensor, the Einstein equations 
(\ref{solnset1})--(\ref{solnset3}), with $\Lambda_{phys}=0$ are
reduced to the following two equations for the metric factors outside
the 3-brane source
\begin{eqnarray}
\label{pfsoln1}
    (n-1)!\frac{Q^2}{\gamma^{n-1}}  &-& \frac{1}{2\gamma}(n-2)(n+2)
    +\frac{\Lambda_D}{M_D^{n+2}} =0~,    \\
\label{pfsoln2}
   c^2 &=& 
   -\frac{1}{2} \frac{\Lambda_D}{M_D^{n+2}}+ \frac{1}{4\gamma}(n-2)^2~.
\end{eqnarray}
We are interested in the solutions of these two equations which 
are exponential, $c^2>0$ and do not change the metric signature, 
$\gamma>0$. Remarkably, solutions to these equations 
exist for which these conditions can be simultaneously satisfied. 
Let us first consider the case $n=2$.
Then the solutions reduce simply to
\begin{eqnarray}
    \frac{Q^2}{\gamma}  &=& -\frac{\Lambda_6}{M_6^4}~, \\
     c^2 &=& -\frac{1}{2} \frac{\Lambda_6}{M_6^4}~.
\end{eqnarray}
Thus, for $\Lambda_6 < 0$ we see that there is solution satisfying 
(\ref{pfansatz}). This solution is different from the local string
defect~\cite{gs}. In this case the brane tension components satisfy
\begin{equation}
    \mu_0= \mu_\theta + (1-\frac{Q}{2\sqrt{2}})A^2 M_6^4~,
\end{equation}
where $\mu_\rho$ remains undetermined.
This  reduces to the condition satisfied by the local string 
solution when $Q=0$.

Next we consider the case $n=3$. Only the `+' solution to the quadratic 
equation (\ref{pfsoln1}) gives rise to 
a solution with both $c^2>0$ and $\gamma > 0$. This solution can be written 
in the form
\begin{eqnarray}
    \frac{Q^2}{\gamma}  &=& \frac{1}{4}\left[ \frac{5}{2} + 
     \sqrt{(\frac{5}{2})^2 - 8 Q^2\frac{\Lambda_7}{M_7^5}}\right]~, \\
   Q^2 c^2 &=& \frac{1}{16} \left[ \frac{5}{2} - 8 Q^2
      \frac{\Lambda_7}{M_7^5}
     +\sqrt{(\frac{5}{2})^2 - 8 Q^2\frac{\Lambda_7}{M_7^5}}\right]~.
\end{eqnarray}
When $Q^2\Lambda_7/M_7^5 < 25/32 $ we obtain real solutions
which are plotted in Figure~\ref{fig:soln3}.
\begin{figure}[ht]
\centerline{ \epsfxsize 4.0 truein \epsfbox {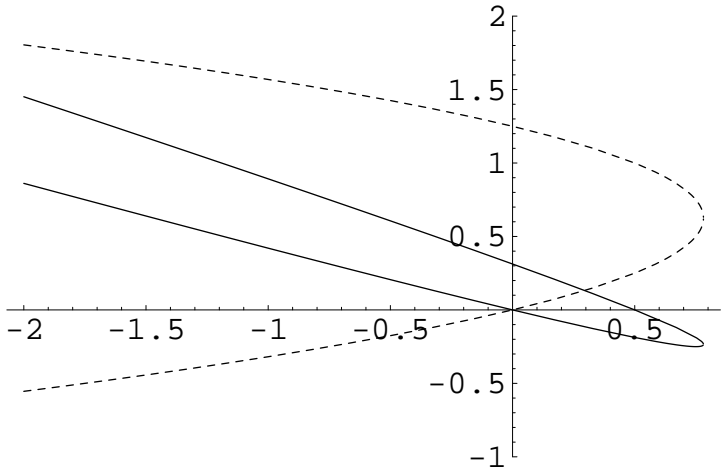}}
\caption{\it The $n=3$ solution for $Q^2/\gamma$ (dashed line)
and $Q^2 c^2$ (solid line), as a function of $Q^2 \Lambda_7/M_7^5$.
Only the branches with $c^2 >0$ and $\gamma>0$ lead to solutions
that localize gravity.}
\label{fig:soln3}
\end{figure}
In fact requiring $c^2>0$ we
see that there are solutions not only for $\Lambda_7< 0$, but also 
for $\Lambda_7 \geq0$, provided that $Q^2\Lambda_7/M_7^5 < 1/2$. 
Thus, the bulk cosmological constant does not need to
be negative in order to localize gravity.

Similarly, solutions for which $c^2>0$ and $\gamma>0$ exist for values
of $n>3$. Again, we find that solutions with
both positive, zero and negative bulk cosmological constants exist.
In general for these type of solutions, 
the 3-brane tension components satisfy
\begin{equation}
     \mu_0= \mu_\theta -\frac{c}{2}\gamma^{(n-1)/2} M_D^{n+2}~,
\end{equation}
where $\mu_\rho$ remains undetermined.

The nice property of these solutions (\ref{pfansatz})
is that since $\gamma$ is a constant
the Ricci scalar does not blow up at any point in the transverse space. 
In particular for the $n=2$ solution the Ricci scalar is
$R=-5/2 \Lambda_6/M_6^4$, while for $n=3$ it is
\begin{equation}
     R= -\frac{1}{2Q^2}\left[ \frac{35}{16} + Q^2\frac{\Lambda_7}{M_7^5}
     + \frac{7}{8}
     \sqrt{(\frac{5}{2})^2 - 8 Q^2\frac{\Lambda_7}{M_7^5}}\right]~.
\end{equation}
The space is again a constant curvature space, but it is not
necessarily anti-deSitter. This is also confirmed by checking
the higher-order curvature invariants $R_{AB}R^{AB}$ and $R_{ABCD}R^{ABCD}$.

Finally, we see from (\ref{diffop}) that
the equation of motion for the spin-2 radial modes using the solution
(\ref{pfansatz}) is qualitatively similar to the 5d case. The constant
$\gamma$ factor effectively plays no role in the localization of gravity.
Thus, the corrections to Newton's law will be suppressed by $1/r^2$
for all solutions $n\geq 3$. This is easy to understand since the
geometry of the extra dimensions is simply $R_+\times S_{n-1}$, as in the
global defect case, and there is just one non-compact
dimension for all $n\geq 3$. In the special case when 
$\Lambda_D=0$, the metric solution we have found corresponds to the 
near-horizon metric of a class of extremal nondilatonic black 
branes~\cite{gibb}.

Another possible solution for the $p$-form in the bulk includes
\begin{equation}
\label{osoln}
      F_{\mu_1\dots\mu_n} = \epsilon_{\mu_1\dots\mu_n} \kappa(\rho)~,
\end{equation}
where 
\begin{equation}
        \kappa(\rho)= Q \frac{\gamma^{(n-1)/2}}{\sigma^2}
(\sin\theta_{n-1})^{(n-2)}\dots \sin\theta_2~.
\end{equation}
In this case the contribution to the stress-energy tensor is
\begin{eqnarray}
      T^\mu_\nu &=& \frac{n!}{4}\frac{Q^2}{\sigma^4} \delta^\mu_\nu~, \\
      T^\rho_\rho &=& T^\theta_\theta = -\frac{n!}{4}\frac{Q^2}{\sigma^4}~.
\end{eqnarray}
This contribution does not appear to make the solution of the 
Einstein equations any easier.

The above two $p$-form solutions have only included components 
in the transverse space. If we also require 
the $p$-form field to transform nontrivially under the 3-brane 
coordinates, then we can have 
\begin{equation}
     F_{0123 \theta_1\dots \theta_{n-1}} = 
     Q(\sin\theta_{n-1})^{(n-2)}\dots \sin\theta_2~,
\end{equation}
where all other components of $F$ are zero. The components
of the stress-energy tensor for this field configuration are
\begin{eqnarray}
      T^\mu_\nu &=&
      \frac{(n+3)!}{4}\frac{Q^2}{\sigma^4\gamma^{n-1}}
      \delta^\mu_\nu \\ T^\rho_\rho &=& -T^\theta_\theta =
      -\frac{(n+3)!}{4}\frac{Q^2} {\sigma^4\gamma^{n-1}} ~.
\end{eqnarray}
Again, there is no simple solution of the Einstein equations
with the inclusion of this contribution.

Finally, one can also generalize the solution (\ref{osoln})
\begin{equation}
      F_{\mu\nu\alpha\beta a_1\dots a_n} =
        \epsilon_{\mu\nu\alpha\beta a_1\dots a_n}  \kappa(\rho)~,
\end{equation}
where 
\begin{equation}
     \kappa(\rho)= Q \sigma^2 \gamma^{(n-1)/2}
(\sin\theta_{n-1})^{(n-2)}\dots \sin\theta_2~,
\end{equation}
and the stress energy tensor
\begin{equation}
      T^M_N = \frac{(n+4)!}{4} Q^2 \delta^M_N~,
\end{equation}
is a constant. Thus, we see that the addition of a
$n+4$-form field is equivalent to adding a bulk cosmological
constant.

\section{Conclusion} 

We have seen that higher-dimensional solutions exist which can
localize  gravity to the 3-brane. The generalization of the
exponential solution found  in Ref.~\cite{rusu,gs}, only exists when a
scalar field with ``hedgehog'' configuration is added to the bulk. In
this case the transverse space no longer  has constant curvature, and
furthermore it develops a singularity at $\rho=\infty$. The
corrections to Newton's law on the 3-brane are suppressed by $1/(c
r)^{n+1}$. If, however $\gamma$ is a constant then regular solutions 
with an exponential warp factor do exist.

However, if instead a $n-1$-form field configuration is added in the
bulk, which generalizes the magnetic field of a monopole, then
solutions which localize gravity can be found for positive, zero
and negative bulk cosmological constant. In this case, the
transverse space has constant curvature but is not an anti-deSitter
space. The corrections to Newton's law have the same form as in the
original model~\cite{rusu}. Furthermore, the addition of an $n+4$-form
field in the bulk is equivalent to adding a bulk cosmological constant.

Given that $p$-form fields are generic in string theories, it would
be interesting to study whether the exponential solutions that we
have found here can be realized in an effective supergravity theory
(see also \cite{emil}).
It is encouraging that the embedding of dilatonic ``global cosmic
strings'' in string theory has recently  been considered
in~\cite{bhm,cpt}.

\noindent {\it Acknowledgments:} We wish to thank E. Poppitz, 
S. Randjbar-Daemi, P. Tinyakov and V. Rubakov for helpful discussions. 
This work was supported by the FNRS, contract no. 21-55560.98.


\begin{thebibliography}{99}


\bibitem{RS2}
V.~A.~Rubakov and M.~E.~Shaposhnikov,
Phys.\ Lett.\  {\bf B125} (1983) 136.

\bibitem{RS1}
V.~A.~Rubakov and M.~E.~Shaposhnikov,
Phys.\ Lett.\  {\bf B125} (1983) 139.

\bibitem{akama}
K.~Akama, in {\it Proceedings of the Symposium on Gauge Theory and 
Gravitation}, Nara, Japan, eds. K.~Kikkawa, N.~Nakanishi and H.~Nariai 
(Springer-Verlag, 1983), [hep-th/0001113].

\bibitem{visser}
M.~Visser,
Phys.\ Lett.\  {\bf B159} (1985) 22
[hep-th/9910093].

\bibitem{anton}
I.~Antoniadis,
Phys.\ Lett.\  {\bf B246} (1990) 377.

\bibitem{add}
N.~Arkani-Hamed, S.~Dimopoulos and G.~Dvali,
Phys.\ Lett.\  {\bf B429} (1998) 263
[hep-ph/9803315];
I.~Antoniadis, N.~Arkani-Hamed, S.~Dimopoulos and G.~Dvali,
Phys.\ Lett.\  {\bf B436} (1998) 257
[hep-ph/9804398].

\bibitem{rusu}
L.~Randall and R.~Sundrum,
Phys.\ Rev.\ Lett.\  {\bf 83} (1999) 4690
[hep-th/9906064].

\bibitem{KK}
{\it Modern Kaluza-Klein Theories}, 
eds. T.~Appelquist, A.~Chodos and P.~G.~Freund,
(Addison-Wesley, 1987).

\bibitem{branes}
J.~Polchinski,
Phys.\ Rev.\ Lett.\  {\bf 75} (1995) 4724
[hep-th/9510017].

\bibitem{greg}
R.~Gregory, V.~A.~Rubakov and S.~M.~Sibiryakov,
Phys.\ Rev.\ Lett.\  {\bf 84} (2000) 5928
[hep-th/0002072]

\bibitem{gs}
T.~Gherghetta and M.~Shaposhnikov,
hep-th/0004014.

\bibitem{cp}
A.~Chodos and E.~Poppitz,
Phys.\ Lett.\  {\bf B471} (1999) 119
[hep-th/9909199].

\bibitem{ran} 
S.~Randjbar-Daemi and C.~Wetterich,
Phys.\ Lett.\  {\bf B166} (1986) 65.

\bibitem{ck}
A.~G.~Cohen and D.~B.~Kaplan,
Phys.\ Lett.\  {\bf B470} (1999) 52
[hep-th/9910132].

\bibitem{rg}
R.~Gregory,
Phys.\ Rev.\ Lett.\  {\bf 84} (2000) 2564
[hep-th/9911015].

\bibitem{ov}
I.~Olasagasti and A.~Vilenkin,
hep-th/0003300.

\bibitem{gib}
G.~W.~Gibbons and D.~L.~Wiltshire,
Nucl.\ Phys.\  {\bf B287} (1987) 717.

\bibitem{fiu}
V.~P.~Frolov, W.~Israel and W.~G.~Unruh,
Phys.\ Rev.\  {\bf D39} (1989) 1084.

\bibitem{dvali}
G.~Dvali,
hep-th/0004057.

\bibitem{oda}
I.~Oda,
hep-th/0006203.

\bibitem{luty}
J.~Chen, M.~A.~Luty and E.~Ponton,
hep-th/0003067.

\bibitem{cpt}
A.~Chodos, E.~Poppitz and D.~Tsimpis,
hep-th/0006093.

\bibitem{cehs}
C.~Csaki, J.~Erlich, T.~J.~Hollowood and Y.~Shirman,
hep-th/0001033.

\bibitem{bend}
J.~Garriga and T.~Tanaka,
hep-th/9911055; 
S.~B.~Giddings, E.~Katz and L.~Randall,
JHEP {\bf 0003} (2000) 023
[hep-th/0002091].

\bibitem{lt}
G.~V.~Lavrelashvili and P.~G.~Tinyakov,
Sov.\ J.\ Nucl.\ Phys. {\bf 41} (1985) 172.

\bibitem{gibb}
G.~W.~Gibbons, G.~T.~Horowitz and P.~K.~Townsend,
Class.\ Quant.\ Grav.\  {\bf 12} (1995) 297
[hep-th/9410073].

\bibitem{emil}
E.~Dudas and J.~Mourad,
hep-th/0004165.

\bibitem{bhm}
P.~Berglund, T.~Hubsch and D.~Minic,
hep-th/0005162.

\end{thebibliography}
\end{document}